\documentclass{article}
\usepackage{amsmath}

\input{tcilatex}

\begin{document}

\title{Inflation conditions for non-BPS D-branes }
\author{Pawel Gusin \\
University of Silesia, Institute of Physics, ul. Uniwersytecka 4, \\
PL-40007 Katowice, Poland}
\date{}
\maketitle

\begin{abstract}
We investigate the effective action for a non-BPS brane in a time-depending
embedding. This action is considered as the action for tachyon and embedding
coupled to the brane gravity. We derive the slow roll parameters from this
model.

\begin{description}
\item[PACS] 11.25.Wx ; 98.80.-k
\end{description}
\end{abstract}

\section{Introduction}

The observation of Cosmic Microwave Bacground (CMB) has provided good
evidence that the universe, it being described by the
Friedman-Robertson-Walker (FRW) metric, underwent a period of acceleration
in its early times. The data from Type Ia super-novae pointed out that the
universe has accelerated very recently and remains in this state up till
now. To describe these phenomena in the string theory approach makes one of
the most important theoretical challenges. In seeking a description of these
phenomena one may pass to an effective field theory in the low energetic
approximation. The result then is a supergravity theory in 10 dimensions. In
order to obtain 3-spatial dimensions one shall either construct a
spontaneous compactification scenario or postulate that universe is a type
of a 3-brane. One of the most popular recent approaches to the inflation
problem is to use an open string tachyon on a non-BPS brane as an inflaton
[1]. The non-BPS states are then realized as the bounded states of a
brane-antibrane system with tachyon condensation [2]. In this approach
(scenario) the inflaton potential should ,in principle, be computed directly
by substituting the complete superpotential into the supergravity
F-potential. Then the break of supersymmetry in the brane-antibrane system
leads to a subtle problem. The problem is that the exponential tachyon
potential cannot produce the last 60 e-folds [3]. In the other scenario the
role of inflaton is played by the separation between D-branes\emph{\ }[4,
5]. Both of the scenarios above are accommodated in the form of a hybrid
inflation where the tachyoinc open string fluctuations end inflation [6]. In
this paper we shall study the inflation conditions in the system with a
tachyon field. This system corresponds to a non-BPS brane and is described
by the DBI-like action. This system is embedded in the background produced
by BPS branes. The effective action for a non-BPS brane consists of the
Hilbert-Einstein action and the DBI-like action. The form of the slow roll
parameters is obtained from this effective action.

\section{Non-BPS Dp-branes}

A $N$ coincident BPS Dk-branes produced a background in which metric $G_{MN}$%
\ a dilaton $\phi $ and the RR potential $\widetilde{A}_{\left( k+1\right) }$
are given by [7]:

\begin{equation}
G_{MN}dX^{M}dX^{N}=\lambda \eta _{\mu \nu }dX^{\mu }dX^{\nu }+\lambda
^{-1}\left( dr^{2}+g_{mn}dX^{m}dX^{n}\right) ,  \tag{2.1}
\end{equation}%
\begin{equation}
e^{-\phi }=\lambda ^{\left( 3-k\right) /2},  \tag{2.2}
\end{equation}%
\begin{equation}
\widetilde{A}_{\left( k+1\right) }=\lambda ^{2}dt\wedge dX^{1}\wedge
...\wedge dX^{k},  \tag{2.3}
\end{equation}%
where the warp factor $\lambda $ is $\lambda =\left[ H_{k}\left( r\right) %
\right] ^{-1/2}$ and $H_{k}\left( r\right) $ is a harmonic function. In the
warped compactifications the factor $\lambda $ is constrained [8]. These BPS
Dk-branes warp (8-k)-dimensional manifold $Y$ with a metric $g_{mn}$.

Let us consider a non-BPS Dp-brane (with $p<k$) which is embedded in the
background descirbed above. The action for this non-BPS brane is [9]:%
\begin{equation}
S=-T_{p}\int d^{p+1}\xi \widetilde{v}\left( T\right) e^{-\phi }\sqrt{-\det
(\gamma _{\mu \nu }+2\pi \alpha ^{\prime }F_{\mu \nu }+B_{\mu \nu }+\partial
_{\mu }T\partial _{\nu }T)}+T_{p}\int v\left( T\right) dT\wedge X^{\ast }%
\widetilde{A}_{\left( k+1\right) },  \tag{2.4}
\end{equation}%
where $T$ is a tachyon field with a potential $\widetilde{v}$.

For embedding in the following form:%
\begin{equation}
X^{M}\left( t,\xi ^{1},...,\xi ^{p}\right) =\left( t,\xi ^{1},...,\xi
^{p},r\left( t,\xi ^{1},...,\xi ^{p}\right) ,\theta ^{1},...,\theta
^{8-p}\right) ,  \tag{2.5}
\end{equation}%
the action (2.4) (for $F=B=0$) takes on the form:%
\begin{equation}
S=-T_{p}\int d^{p+1}\xi \widetilde{v}\left( T\right) e^{-\phi }\sqrt{-\det
(\lambda \eta _{\mu \nu }+\lambda ^{-1}\partial _{\mu }r\partial _{\nu
}r+\partial _{\mu }T\partial _{\nu }T)}+T_{p}\int v\left( T\right) dT\wedge
X^{\ast }\widetilde{A}_{\left( k+1\right) }  \tag{2.6}
\end{equation}%
and the induced metric $\gamma _{\mu \nu }$ is:%
\begin{equation}
\gamma _{\mu \nu }=\lambda \eta _{\mu \nu }+\lambda ^{-1}\partial _{\mu
}r\partial _{\nu }r.  \tag{2.6a}
\end{equation}%
The DBI-like part can be rewritten as follows:%
\begin{equation}
S=-\int d^{p+1}\xi v\left( T\right) \lambda ^{\left( 4+p-k\right) /2}\sqrt{%
\det (I+\eta ^{-1}S)},  \tag{2.7}
\end{equation}%
where the matrix $S$ has entries:%
\begin{equation}
S_{\mu \nu }=\lambda ^{-2}\partial _{\mu }r\partial _{\nu }r+\lambda
^{-1}\partial _{\mu }T\partial _{\nu }T  \tag{2.8}
\end{equation}%
and $v\left( T\right) =T_{p}\widetilde{v}\left( T\right) $. We restrict
ourselves to the case when the tachyon $T$ and the field $r$ depend only on
time $t.$ Thus the action takes on the form [10]:%
\begin{equation}
S=-\int d^{p+1}\xi v\left( T\right) \lambda ^{\left( 4+p-k\right) /2}\sqrt{%
1-\lambda ^{-2}\overset{\cdot }{r}^{2}-\lambda ^{-1}\overset{\cdot }{T}^{2}}%
+T_{p}\int v\left( T\right) dT\wedge X^{\ast }\widetilde{A}_{\left(
k+1\right) }.  \tag{2.9}
\end{equation}

The DBI-like action (2.9) is appropriate for distances $r$ larger then the
fundamental string length $l_{s}$ between a Dp-brane and a background
k-brane. Otherwise one should replace this action with\emph{\ }the action of
a complex scalar tachyon field with a potential. This potential was
calculated in [11] for p=3 and k=5.

\section{Inflation and slow-roll parameters}

We investigate cosmological consequences of the action (2.9). This action is
considered as the action for the fields $r,T$ which are coupled to the
Einstein gravity on the world-volume of the brane. The action for the
tachyonic field only is considered in [12]. We also restrict the dimension $%
p $ of a non-BPS brane to 3. Thus the effective action for a non-BPS
D3-brane is given by:%
\begin{equation}
S_{eff}=\int d^{4}x\frac{m_{P}^{2}}{2}\sqrt{-\gamma }R+S[r,T],  \tag{3.1}
\end{equation}%
where the 4-dimensional Planck mass $m_{P}$ is equal to $\left( 8\pi
G\right) ^{-1/2}$ and the scalar curvature $R$ is obtained from the metric
(2.6a). The action $S[r,T]$ is given by (2.9). In the case when $r$ is
homogenous and depends on time $t$ the induced metric on the worldvolume\
has the form:%
\begin{equation}
ds^{2}=-\sigma dt^{2}+\lambda \delta _{mn}dx^{m}dx^{n},  \tag{3.2}
\end{equation}%
where:%
\begin{equation}
\sigma =\lambda \left( r\right) -\frac{\overset{\cdot }{r}^{2}}{\lambda
\left( r\right) }.  \tag{3.3}
\end{equation}%
The Lagrangian for fields $r$ and $T$ is obtained from (2.9) and has the
form: $L=v\left( T\right) e^{-\Phi }\sqrt{1-\overset{\cdot }{T}^{2}/\sigma }$
where $\Phi =\phi -\frac{3}{2}\ln \lambda -\frac{1}{2}\ln \sigma $. The
energy-momentum tensor for the above system is:%
\begin{equation}
T_{00}=\frac{\sigma ve^{-\phi }}{\sqrt{1-\overset{\cdot }{T}^{2}/\sigma }}, 
\tag{3.4}
\end{equation}%
\begin{equation}
T_{mn}=-\lambda v\left( T\right) e^{-\phi }\left( 1-\overset{\cdot }{T}%
^{2}/\sigma \right) ^{1/2}\delta _{mn}.  \tag{3.5}
\end{equation}%
Thus the field equations $R_{\mu \nu }-\frac{1}{2}\gamma _{\mu \nu }R=8\pi
GT_{\mu \nu }$ takes on the form:%
\begin{equation}
H^{2}+\frac{1}{2}\frac{\overset{\cdot }{\sigma }}{\sigma }H\left( \frac{%
\sigma }{a^{2}}-1\right) =8\pi G\frac{\sigma ve^{-\phi }}{3\sqrt{1-\overset{%
\cdot }{T}^{2}/\sigma }},  \tag{3.6}
\end{equation}%
\begin{equation}
2\frac{\overset{\cdot \cdot }{a}}{a}+H^{2}-\frac{\overset{\cdot }{\sigma }}{%
\sigma }H=8\pi G\sigma ve^{-\phi }\left( 1-\overset{\cdot }{T}^{2}/\sigma
\right) ^{1/2},  \tag{3.7}
\end{equation}%
where $a^{2}=\lambda $ and the Hubble parameter $H$ is given by $H=\overset{%
\cdot }{a}/a$. The equation of motion for $T$ is obtained from the
Lagrangian $L$: 
\begin{equation}
\frac{\overset{\cdot \cdot }{T}}{1-\overset{\cdot }{T}^{2}/\sigma }+\left(
6-k\right) H\overset{\cdot }{T}+\frac{v^{\prime }}{v}\sigma -\frac{1}{2}%
\frac{\overset{\cdot }{T}}{1-\overset{\cdot }{T}^{2}/\sigma }\frac{\overset{%
\cdot }{\sigma }}{\sigma }=0.  \tag{3.8}
\end{equation}%
For $\sigma =1$ the field $\Phi $ is related to the warp factor $\lambda $
as follows $e^{-\Phi }=\lambda ^{\beta +1/2}$. Let $\beta +1/2=\left(
3-k\right) /2$. Thus the equations (3.6) and (3.7) are reduced to the form
(note that $e^{-\phi }=a^{2\beta +1}$):%
\begin{equation}
H^{2}=8\pi G\frac{va^{2\beta +1}}{3\sqrt{1-\overset{\cdot }{T}^{2}}}, 
\tag{3.9}
\end{equation}%
\begin{equation}
\frac{\overset{\cdot \cdot }{a}}{a}=8\pi G\frac{va^{2\beta +1}}{3\sqrt{1-%
\overset{\cdot }{T}^{2}}}\left( 1-3\overset{\cdot }{T}^{2}/2\right) . 
\tag{3.10}
\end{equation}%
The constraint $\sigma =1$ says that the metric (3.2) is space flat with the
scale factor $a^{2}$. For $\beta =-1/2$ (which corresponds to $k=3$) and $%
\sigma =1$ we obtain the well-known form of the equations. We shall only
consider the case when $\sigma =1$.

In order to get conditions on inflation we use the slow-roll parameters from
[13]. In [14] a similar problem was considered but it did not account for
the dilaton field. These slow-roll parameters are defined as follows:%
\begin{equation}
\varepsilon _{i+1}=\frac{d\ln |\varepsilon _{i}|}{dN},  \tag{3.11}
\end{equation}%
where $\varepsilon _{0}=H_{0}/H$ and $H_{0}$ is the Hubble parameter at some
chosen time. The Hubble parameter is considered here as the function of the
e-foldings number $N$ given by: $N=\int_{t_{init}}^{t_{end}}Hdt$ . The
parameters $\varepsilon _{i}$ as the functions of time $t$ are governed by
the equation:%
\begin{equation}
H\varepsilon _{i}\varepsilon _{i+1}=\overset{\cdot }{\varepsilon }_{i}. 
\tag{3.12}
\end{equation}%
The first two slow-roll parameters have the from :%
\begin{equation}
\varepsilon _{1}=-\frac{1}{H}\frac{dH}{dT}\frac{dT}{dN}=-\left( \beta
+1/2\right) +\left( \beta +2\right) \overset{\cdot }{T}^{2},  \tag{3.13}
\end{equation}%
\begin{equation}
\varepsilon _{2}=\frac{1}{\varepsilon _{1}}\frac{d\varepsilon _{1}}{dT}\frac{%
dT}{dN}=\frac{2\left( \beta +2\right) \overset{\cdot }{T}\overset{\cdot
\cdot }{T}}{\left[ -\left( \beta +1/2\right) +\left( \beta +2\right) \overset%
{\cdot }{T}^{2}\right] H},  \tag{3.14}
\end{equation}%
where we used equations (3.8) (with $\sigma =1$), (3.9) and the relation:$%
dT/dN=\overset{\cdot }{T}/H$. Thus the equation (3.9) as the function of $%
\varepsilon _{1}$ takes on the form:%
\begin{equation}
H^{2}\sqrt{1-\frac{2}{3}\varepsilon _{1}}=\frac{8\pi G}{3\sqrt{3}}va^{2\beta
+1}\sqrt{2\beta +4}.  \tag{3.15}
\end{equation}%
Differentiation of the above equation, with respect to the cosmological time 
$t$ gives (where we used (3.12)):%
\begin{equation}
-2\sqrt{\left( \beta +2\right) \widetilde{\varepsilon }_{1}}\frac{\left[ 1-%
\frac{2}{3}\varepsilon _{1}+\frac{1}{6}\eta \varepsilon _{2}\right] }{%
1-2\varepsilon _{1}/3}=\frac{v^{\prime }}{vH},  \tag{3.16}
\end{equation}%
where: $\widetilde{\varepsilon }_{1}=\varepsilon _{1}+\beta +1/2$ and $\eta
=\varepsilon _{1}/\widetilde{\varepsilon }_{1}$. The second derivative of
(3.15) gives:%
\begin{gather}
\left( 2\varepsilon _{1}-\eta \varepsilon _{2}\right) +\frac{\varepsilon _{2}%
}{3}\left[ 5\varepsilon _{1}-\frac{\eta \left( 3\eta -2\right) }{2}%
\varepsilon _{2}-\eta \varepsilon _{3}\right] \gamma ^{2}+  \notag \\
+4\widetilde{\varepsilon }_{1}\left( 1-\frac{2}{3}\varepsilon _{1}-\frac{1}{6%
}\eta \varepsilon _{2}\right) \left( 1-\frac{2}{3}\varepsilon _{1}+\frac{1}{6%
}\eta \varepsilon _{2}\right) \gamma ^{4}=\frac{v^{\prime \prime }}{\left(
\beta +2\right) vH^{2}},  \tag{3.17}
\end{gather}%
where $\gamma ^{2}=\left( 1-2\varepsilon _{1}/3\right) ^{-1}$ . Up to the
first order in $\varepsilon _{1}$ and $\varepsilon _{2}$ we get:%
\begin{equation}
\varepsilon _{1}=\left( \frac{3}{2}\right) ^{5/2}\frac{m_{Pl}^{2}}{\left(
2+\beta \right) ^{3/2}\left( 1-4\beta \right) }\frac{v^{\prime 2}}{v^{3}}%
e^{\phi }-\frac{3}{2}\frac{1+2\beta }{1-4\beta },  \tag{3.18}
\end{equation}%
\begin{equation}
\eta \varepsilon _{2}=3\sqrt{\frac{3}{2}}\frac{m_{Pl}^{2}}{\left( 2+\beta
\right) ^{3/2}}\left[ \frac{4-\beta }{1-4\beta }\frac{v^{\prime 2}}{v^{3}}-%
\frac{v^{\prime \prime }}{v^{2}}\right] e^{\phi }-6\frac{\left( 1+\beta
\right) \left( 1+2\beta \right) }{1-4\beta },  \tag{3.19}
\end{equation}%
where $e^{\phi }=a^{-2\beta -1}$. From (3.19) we get the second parameter $%
\varepsilon _{2}$ expressed by $\varepsilon _{1}$:%
\begin{equation}
\varepsilon _{2}=\frac{4\left( 4-\beta \right) }{3}\varepsilon _{1}-3s\frac{%
v^{\prime \prime }}{v^{2}}+\left( 1+2\beta \right) \left[ \frac{2}{3}\left(
7-\beta \right) +\frac{2\left( 1+2\beta \right) -3sv^{\prime \prime }/v^{2}}{%
2\varepsilon _{1}}\right] ,  \tag{3.20}
\end{equation}%
where $s=\sqrt{3/2}m_{Pl}^{2}\left( 2+\beta \right) ^{-3/2}e^{\phi }$ . The
inflation takes place if $0<\varepsilon _{1}<1$. The number of e-foldings,
expressed in terms of the tachyon field $T$ and the dilaton field $\phi $ is:%
\begin{gather}
N=\int_{T}^{T_{end}}\frac{H}{\overset{\cdot }{T}}dT=-\left( \frac{2}{3}%
\right) ^{3/2}\frac{\left( 2+\beta \right) ^{3/2}\left( 1+16\beta +4\beta
^{2}\right) }{2m_{Pl}^{2}\left( 1-4\beta \right) }\int_{T_{end}}^{T}\frac{%
v^{2}}{v^{\prime }}e^{-\phi }dT+  \notag \\
+\frac{1}{3}\int_{T_{end}}^{T}\left( \frac{11-2\beta }{2\left( 1-4\beta
\right) }\frac{v^{\prime }}{v}-\frac{v^{\prime \prime }}{v^{\prime }}\right)
dT.  \tag{3.21}
\end{gather}
Since the dimension of the manifold $Y$ is $8-k$ (see eqs.(2.1)-(2.3)), the
case $\beta =-1/2$ corresponds to the background produced by the D-branes
which are warping $5$ dimensional manifold. In this case the parameters $%
\varepsilon _{1}$ and $\varepsilon _{2}$ become\ the standard parameters
considered in the tachyon inflation.

\section{Conclusions}

In this paper we considered gravity on the non-BPS D3-brane. The action for
this system consists of the Einstein-Hilbert action and the DBI-like action
for a D3-brane. From this model we derived the slow-roll parameters. These
parameters depend on the potential $v$, the dilaton field $\phi $ and the
dimension of a manifold on which the background D-branes are warped. For the
given potential $v$ and the given background one can compute these
parameters as the functions of $T$ and $r$. The field $r$ is obtained from
the constraint: $\sigma =1$ (eq.(3.3)) on a D3-brane. The inflation is ended
for the fields $T$ and $r$ if $\varepsilon _{1}\left( T_{end},r_{end}\right)
=1$. From $\varepsilon _{1}$ and $\varepsilon _{2}$ one can calculate the
observable parameters as the functions of $T$ and $r$. In case when $\beta
=-1/2$ we get the well-known parameters for the tachyon inflation.

\section{References}

[1] G. W. Gibbons, \textit{Cosmological Evolution of the Rolling Tachyon},
Phys. Lett. \textbf{B537} (2002) 1, hep-th/0204008.

[2] A.Sen, \textit{Stable Non-BPS Bound States Of BPS D-branes},
hep-th/9805019,

[3] L. Kofman, A. Linde, \textit{Problems with Tachyon Inflation}, JHEP 
\textbf{0207}, 004 (2002), hep-th/0205121;

[4] Kyae and Q. Shafi, \textit{Branes and Inflationary Cosmology}, Phys.
Lett. \textbf{B526 }(2002) 379, hep-ph/0111101.

[5] T. Banks, L. Susskind, \textit{Brane-antibrane forces}, hep-th/9511194.

[6] C. P. Burgess, M. Majumdar, D. Nolte, F. Quevedo, G. Rajesh and R. J.
Zhang, \textit{The Inflationary Brane-Antibrane Universe}, JHEP \textbf{07}
(2001) 047, hep-th/0105204.

[7] G. W. Gibbons, K. Maeda, Nucl. Phys. \textbf{B298 }(1988) 741; D.
Garfinkle, G. T. Horowitz, A. Strominger, Phys. Rev. \textbf{D43} (1991)
3140.

[8] S. B. Giddings, S. Kachru, J. Polchinski, \textit{Hierarchies from
Fluxes in String Compactifications}, Phys. Rev. D\textbf{66} (2002) 106006,
hep-th/0105097.

[9] A. Sen, \textit{Supersymmetric world-volume action for non-BPS D-branes}%
, JHEP\textbf{\ 9910} (1999) 008, hep-th/9909062; M. R. Garousi, \textit{%
Tachyon couplings on non-BPS D-branes and Dirac-Born-Infeld action}, Nucl.
Phys. \textbf{B584}, 284 (2000), hep-th/0003122.

[10] J. Kluso\v{n}, \textit{Non-BPS Dp-brane in Dk-Brane Background}, JHEP 
\textbf{0503} (2005) 044, hep-th/0501010.

[11] E. Gava, K. S. Narain, M. H. Sarmadi, Nucl. Phys. \textbf{B504}, 214
(1997).

[12] G. Gibbons, Phys. Lett. B 537 (2002) 1; G. Gibbons, Class. Quant. Grav.
20 (2003) S321-S346; M. Fairbairn and M. H. Tytgat, Phys. Lett. B 546 (2002)
1;

[13] D. J. Schwarz, C. A. Terrero-Escalante, A. A. Garcia, \textit{Higher
order corrections to primordial spectra from cosmological inflation,}\
Phys.Lett. B517 (2001) 243, astro-ph/0106020.

[14]D.A.Steer, F.Vernizzi, \textit{Tachyon inflation: tests and comparison
with single scalar field inflation}, Phys.Rev. D70 (2004) 043527,
hep-th/0310139.

\end{document}